\documentclass[journal=jacsat,manuscript=article]{achemso}
\usepackage{amsmath}
\usepackage{graphicx}
\usepackage{wasysym}
\usepackage{listings}
\usepackage{color}
\usepackage{comment}

\begin{tocentry}
\vfill
\includegraphics[width=\textwidth]{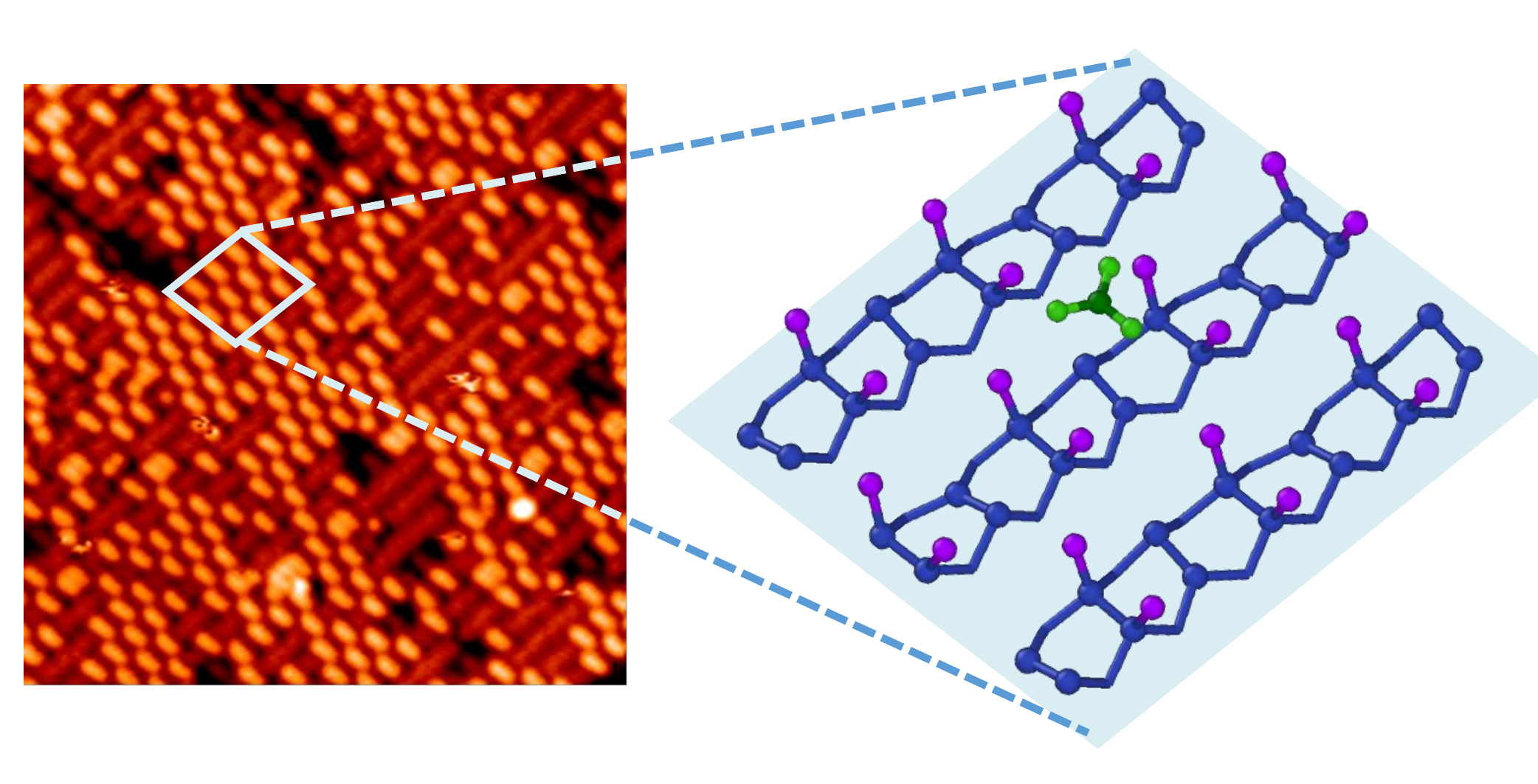}
\vfill
\end{tocentry}

\title{Dopant Precursor Adsorption into Single-Dimer Windows: Towards Guided Self-Assembly of Dopant Arrays on Si(100)}

\author{Matthew S. Radue}
\affiliation{Laboratory for Physical Sciences, 8050 Greenmead Drive, College Park, MD 20740, USA.}
\author{Yifei Mo}
\affiliation{Department of Materials Science and Engineering, University of Maryland, College Park, MD 20742, USA.}
\author{R.E. Butera}
\email{rbutera@lps.umd.edu}
\affiliation{Laboratory for Physical Sciences, 8050 Greenmead Drive, College Park, MD 20740, USA.}
\date{\today}
\begin{document}
\begin{abstract}

Atomically precise dopant arrays in Si are being pursued for solid-state quantum computing applications.  We propose a guided self-assembly process to produce atomically precise arrays of single dopant atoms in lieu of lithographic patterning.  We leverage the self-assembled c(4x2) structure formed on Br- and I-Si(100) and investigate molecular precursor adsorption into the generated array of single-dimer window (SDW) adsorption sites with density functional theory (DFT).  The adsorption of several technologically relevant dopant precursors (PH\textsubscript{3}, BCl\textsubscript{3}, AlCl\textsubscript{3}, GaCl\textsubscript{3}) into SDWs formed with various resists (H, Cl, Br, I) are explored to identify the effects of steric interactions. PH\textsubscript{3} adsorbed without barrier on all resists studied, while BCl\textsubscript{3} exhibited the largest adsorption barrier, 0.34 eV, with an I resist.   Dense arrays of AlCl\textsubscript{3} were found to form within experimentally realizable conditions demonstrating the potential for the proposed use of guided self-assembly for atomically precise fabrication of dopant-based devices.

\end{abstract}

\maketitle

\section{Introduction} \label{introduction}

The ability to place single atoms in atomically precise arrays within solid-state materials has been the focus of intense research for quantum computing, quantum simulation, and spintronics applications \cite{2DQMeta, kane1998, Zhou14378, Khajetoorians2019}.  For dopant atoms in silicon, the precision placement of single dopants with tens-of-nanometers separation would fulfill the challenging specification of envisioned spin-based quantum computers coupled through the exchange \cite{kane1998} or dipole interactions \cite{Salfi:2016}. Placing dopants with a few-atomic-site separation could promote superconductivity in ultra-doped Si \cite{Shim:2014, Blase:2009}, as this would give a high concentration of dopants with still a large enough separation to prevent deactivating defects.\cite{mueller2004,keizer2015}  In addition, it can also be a way to achieve high mobility and reduce statistical variance in electrical properties for devices based on delta-doped Si layers through the engineering of a well-ordered potential.\cite{shinada2005,carter2011}

\begin{figure*}
\centering
\includegraphics[width=\textwidth]{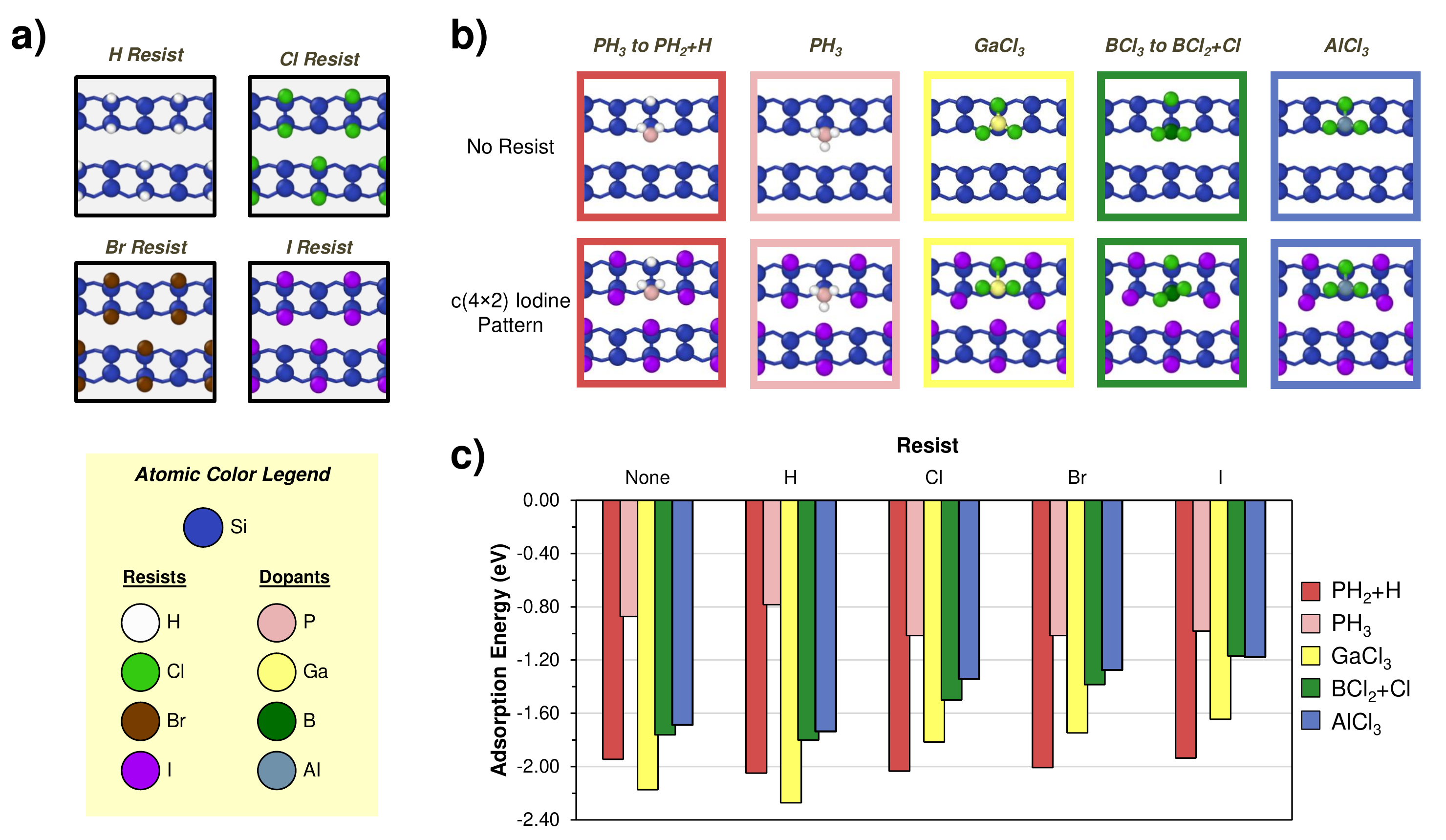}\hfill%
\caption{(a) Models of H-, Cl-, Br-, and I-ADTP surfaces. Below, the atomic color legend for this figure and for all other figures is shown. (b) The adsorbate configurations modeled on Si with no resist (top) and in the windows of the I-ADTP (bottom). (c) Adsorption energies for select adsorbates within a SDW using various resists. The adsorption energy of the adsorbate on bare Si (no resist) is plotted as well. For reference, the data is also tabulated in the Supporting Information.}%
\label{fig_ea}%
\end{figure*}

The current strategy for atomic-precision dopant placement on Si(100) is H depassivation lithography, a scanning tunneling microscopy (STM)-based approach which has been used to successfully place a single P atom within a 3-dimer window.\cite{fuechsle2012} However, fabrication by means of STM is severely limited in its throughput.\cite{randall2018} Moreover, for relevant quantum applications large-scale arrays of dopants need to be placed with the same level of precision.  Here, we propose leveraging naturally occurring, self-assembled resist patterns as an efficient tool for fabricating large-scale atomically precise arrays of dopant atoms, thereby reducing STM demands.  We explore the c(4$\times$2) alternating dimer termination pattern (ADTP), shown in Figure \ref{fig_ea}(a), consisting of an array of alternating adsorbate-terminated Si dimers which has been experimentally observed on both Br- and I-Si(100) \cite{Koji:Bretch, HERRMANN:Br4x2, rioux1995, xu2005}.  It forms as a means to reduce steric repulsion between neighboring adsorbate-terminated dimers\cite{DeWijs:Br4x2, rioux1995,herrmann2000} resulting in a spontaneously formed arrangement of single-dimer windows (SDWs) that have the potential to accommodate the adsorption of a single dopant precursor molecule.

Utilizing the ADTP as a template for atomic-precision dopant placement raises important questions regarding the use of large halogen adatoms as a resist. Only recently has Cl-Si(100) been explored for atomic precision fabrication applications\cite{dwyer2019, pavlova:PH3, pavlova:Cl} and the differences between halogens and hydrogen present a few unknowns. In particular, the increased size of halogen adsorbates relative to hydrogen promote steric hindrances that not only foster the self-assembled resist pattern discussed here, but may also play a role in hindering adsorption into a SDW.

In this study, we use density functional theory (DFT) to explore dopant precursor adsorption into SDWs on H- and X-Si(100)-c(4x2), where X= Cl, Br, and I.  While H- and Cl-Si(100) are not known to form the ADTP pattern, both cases are included to systematically study the role of steric repulsion on adsorption processes.  We model the adsorption of four dopant precursors: PH\textsubscript{3}, BCl\textsubscript{3}, AlCl\textsubscript{3}, and GaCl\textsubscript{3}. PH\textsubscript{3} was selected because of its importance in the field.  The other selected precursors are being explored as candidates for atomically precise acceptor-doping applications\cite{radue2021,dwyer2021}. The findings presented here indicate that SDW adsorption and the formation of well-ordered adsorbate arrays are energetically favorable with minor reaction barriers, supporting the idea of ordering dopants with guided self-assembly.

\section{Methods} \label{methods}

DFT calculations were performed with the VASP program using plane wave basis sets \cite{kresse1993,kresse1994,kresse1996a,kresse1996b}. The Perdew-Burke-Ernzerhof (PBE) exchange-correlation functional of the generalized gradient approximation (GGA) form was selected \cite{perdew1996}. The projector augmented wave (PAW) pseudopotentials were used \cite{blochl1994,kresse1999}.

The Si slab model consisted of two dimer rows and four dimers per row. The ADTP was simulated for each resist type (H, Cl, Br, I). We refer to the ADTP with a given resist X as X-ADTP. Figure \ref{fig_ea}(a) shows each patterned surface. The slab depth was eight Si layers. The bottom Si surface was terminated with H. The two bottom-most Si layers and the terminating H atoms were fixed. The fixed Si atoms were set in conformity to an optimized bulk Si model. A vacuum gap of about 10 \AA~separated the slab from its periodic image. The k-point mesh and energy cutoff were optimized to reach a convergence of 1 meV/atom. The converged mesh was 2$\times$2$\times$1 of the Monkhorst-Pack scheme \cite{monkhorst1976}. A 500 eV energy cutoff was determined to be suitable. Dipole corrections were included.

For a given adsorbate, the adsorption energy $E_{\rm a} $ was calculated as,
\begin{equation} \label{eq1}
E_{\rm a} = E_{\rm slab/adsorbate} - E_{\rm slab} - E_{\rm precursor},
\end{equation}
\sloppy where $E_{\rm slab/adsorbate} $ is the total energy of the adsorbate on the slab including the resist, $E_{\rm slab} $ is the total energy of the slab including the resist but with no adsorbate, and $E_{\rm precursor}$ is the total energy of the isolated precursor. When considering subsequent adsorption steps, the adsorption energy of the second adsorbate was calculated as,
\begin{equation} \label{eq2}
E_{\rm a} = E_{\rm slab/2adsorbates} - E_{\rm slab/1adsorbate} - E_{\rm precursor},
\end{equation}
\sloppy where $E_{\rm slab/2adsorbates} $ is the total energy of both adsorbates on the slab, $E_{\rm slab/1adsorbate} $ is the total energy of the first adsorbate on the slab, and $E_{\rm precursor}$ is the total energy of the isolated precursor associated with the second adsorbate.

To compute the adsorption reaction pathways, the Nudged Elastic Band (NEB) method was used. To obtain additional images along the pathway, nested NEB calculations were performed as desired.

\section{Results and Discussion} \label{results}

\begin{figure*}
\centering
\includegraphics[width=\textwidth]{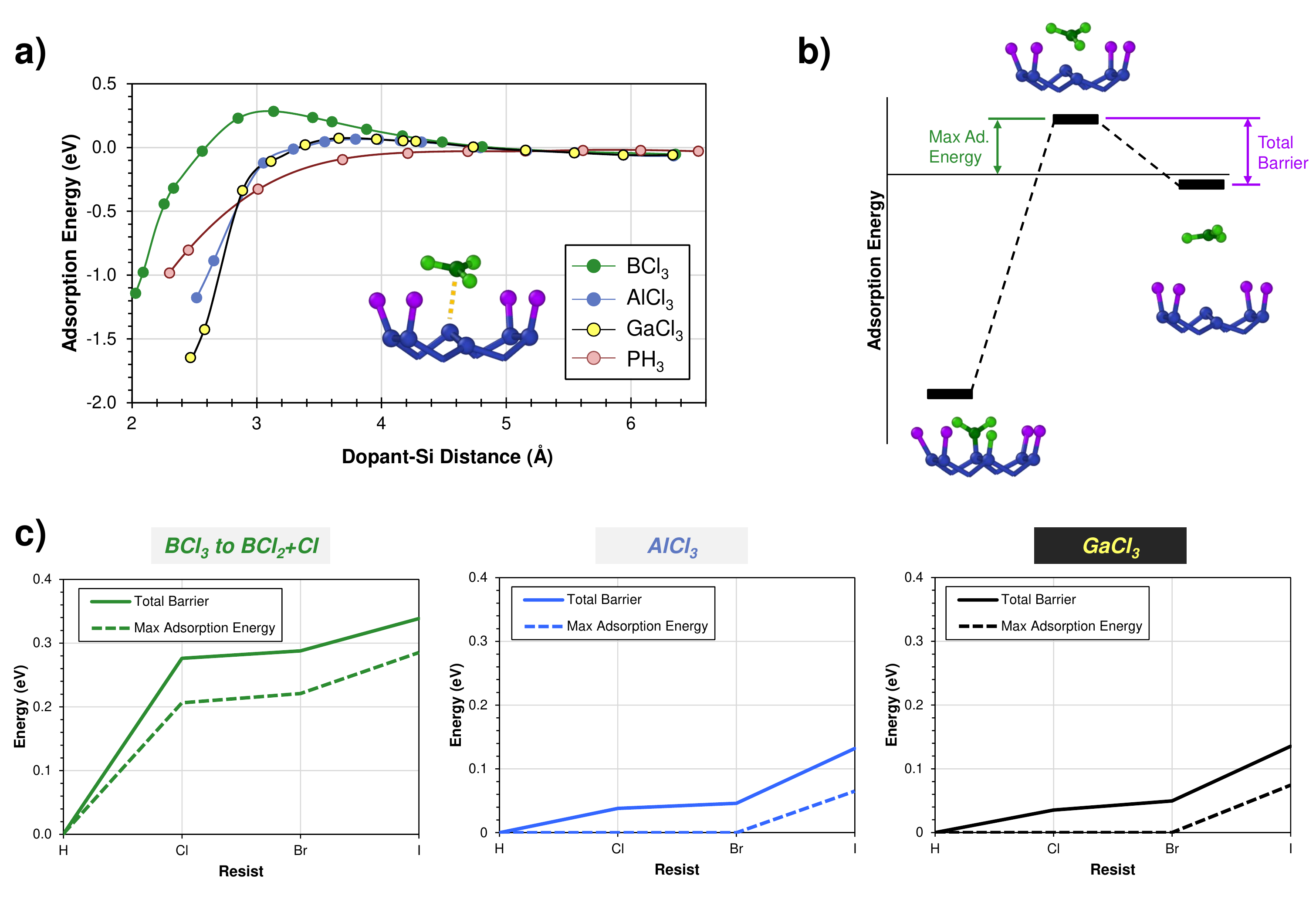}\hfill%
\caption{(a) Reaction pathways for various precursors adsorbing into a SDW on the I-ADTP surface. (b) Schematic plot of adsorption energy as a function of distance from the surface with the maximum adsorption energy and total adsorption barrier defined. (c) Energy barriers for adsorption of various precursors into a SDW of the H-, Cl-, Br-, and I-ADTP. The total energy barriers are also tabulated in the Supporting Information.}%
\label{fig_pathways}%
\end{figure*}

Figure \ref{fig_ea}(b) shows the dopant precursor adsorption configurations we considered. The top row depicts adsorption configurations found on ``bare" Si(100). With the exception of the PH\textsubscript{2}+H structure, the structures pictured are known to be the initial adsorption configurations from previous studies \cite{warschkow2016,ferguson2009, radue2021}. In agreement with Ferguson et al., we find that AlCl\textsubscript{3} and GaCl\textsubscript{3} both form cyclic structures with a Si dimer, while BCl\textsubscript{3} dissociates one Cl to form a BCl\textsubscript{2}+Cl structure \cite{ferguson2009}. While each of the initial acceptor precursor adsorbates occupy an entire Si dimer, PH\textsubscript{3} initially bonds to one Si, and so the PH\textsubscript{2}+H case was also included to better compare with the acceptor adsorbates. Warschkow et al. found that either an on-dimer or inter-dimer PH\textsubscript{2}+H configuration can be reached by surmounting a 0.66 eV or 0.46 eV barrier, respectively, from the undissociated PH\textsubscript{3} adsorbate \cite{warschkow2016}. The on-dimer configuration is of most interest here because it has the potential to fit within a SDW.

The same adsorbate configurations were attempted on the patterned surfaces. The bottom row of Figure \ref{fig_ea}(b) shows the resulting adsorbate configurations for the I resist case. NEB calculations were also performed to explore the energy landscape of rotating PH\textsubscript{3}, the PH\textsubscript{2} fragment, and the BCl\textsubscript{2} fragment, which are shown in the Supporting Information document. From the geometric optimization and NEB calculations, the BCl\textsubscript{2} fragment is found to rotate by about 30 degrees with respect to the BCl\textsubscript{2} orientation on the bare surface, as seen in Figure \ref{fig_ea}(b). In contrast, the phosphine adsorbates do not change in orientation when placed within the window.

The adsorption energies of all the adsorbate-resist combinations studied are shown in Figure \ref{fig_ea}(c). For the acceptor precursors, the favorable energy of adsorption becomes somewhat reduced for the larger resist atoms. This is attributed to the repulsion between the resist atoms and the precursor's Cl atoms. Consider also the positions of the I atoms in close proximity to an acceptor adsorbate (Figure \ref{fig_ea}(b)), where I adatoms are displaced away from the adsorbates. The PH\textsubscript{3}-related adsorbates, however, are mostly insensitive to the resist type due to the small size of PH\textsubscript{3} which can easily fit within the SDW.

\begin{figure*}
\centering
\includegraphics[width=\textwidth]{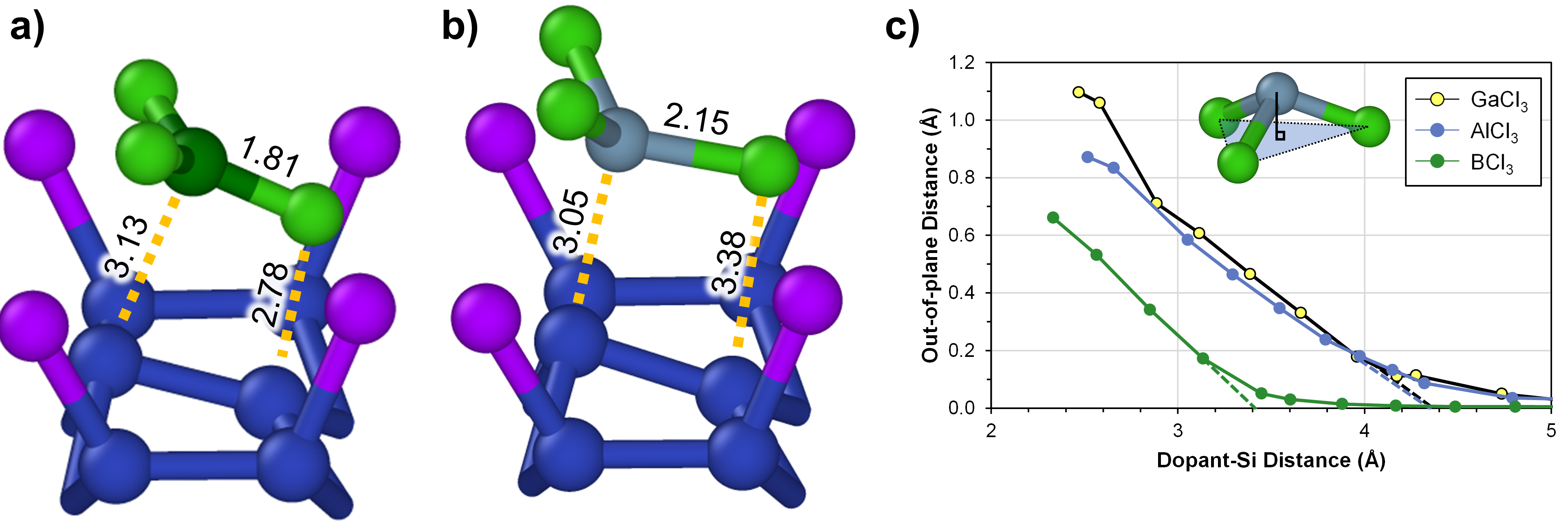}\hfill%
\caption{(a) Approximate transition state for BCl\textsubscript{3} adsorption into a SDW on the I-ADTP surface. (b) AlCl\textsubscript{3} configuration along the adsorption reaction pathway. This configuration is not the maximum energy configuration along the reaction pathway, but it is shown to contrast with the structure in (a). (c) Breaking of the precursor's planar geometry during adsorption onto the I-ADTP surface. The out-of-plane (OOP) distance is the point-plane distance between the central atom and the plane passing through the Cl atoms.}%
\label{fig_geom}%
\end{figure*}

The reaction pathways for adsorption were computed for all precursor-resist combinations. Figure \ref{fig_pathways}(a) shows the adsorption pathways into a SDW for the I-ADTP case. The abscissa indicates the distance between the dopant atom (B, Al, Ga, or P) and the Si atom that directly bonds to it. PH\textsubscript{3} adsorbs without barrier into the window due to the precursor's small size. AlCl\textsubscript{3} and GaCl\textsubscript{3} demonstrate a small barrier (0.13 and 0.14 eV, respectively), but the adsorption barrier for BCl\textsubscript{3} is 0.34 eV, about 2.6 times that of AlCl\textsubscript{3}. To explore the larger energy barrier for BCl\textsubscript{3}, an approximate transition state is shown in Figure \ref{fig_geom}(a). Here, the BCl\textsubscript{3} molecule is planar and close to the surface. This results in marked repulsion between the I and Cl atoms. An image along the AlCl\textsubscript{3} adsorption pathway is shown in Figure \ref{fig_geom}(b) for comparison. This configuration is not the maximum or minimum energy configuration along the reaction pathway. Rather, it is selected to show AlCl\textsubscript{3} at a similar distance from the Si surface as in Figure \ref{fig_geom}(a). In the shown AlCl\textsubscript{3} structure, the Al is able to transition to a tetrahedral geometry, which allows the Cl atoms to better avoid the I atoms. Al is able adopt a tetrahedral geometry sooner because the Al-Si bond is longer than the B-Si bond. For example,  the B-Si bond is 2.02 \AA~for the BCl\textsubscript{2}+Cl adsorbate on bare Si (Figure \ref{fig_ea}(b)); whereas, the Al-Si bond is 2.46 \AA~for the AlCl\textsubscript{3} case. (The GaCl\textsubscript{3} adsorbate has a Ga-Si bond length of 2.42 \AA.) The short B-Si bond means that BCl\textsubscript{3} must get closer to the surface first, coming down in a planar configuration while opposed by the resist atoms, before forming the favorable bond with Si. This is further illustrated in Figure \ref{fig_geom}(c), which tracks the geometry of each acceptor precursor during adsorption into a SDW on the I-ADTP surface. The breaking of the precursor's planar geometry is quantified by the point-plane distance between the central atom (intended dopant) and the plane passing through the Cl atoms, as depicted in Figure \ref{fig_geom}(c). Using the images of the adsorption reaction pathway, we see that AlCl\textsubscript{3} is able to initiate bonding and break out of the planar configuration at a further distance from the surface than BCl\textsubscript{3}. To estimate the onset of out-of-plane (OOP) bending, the OOP distances ranging from $\sim$0.1--0.7 \AA~were linearly fit, and the fit was extrapolated (as shown by the dashed lines in Figure \ref{fig_geom}(c)) to find its x-intercept. According to this estimation, BCl\textsubscript{3} starts bending at a B-Si distance of 3.4 \AA; whereas, AlCl\textsubscript{3} starts bending at an Al-Si distance of 4.3 \AA. The increased barrier for BCl\textsubscript{3} adsorption may be surprising because one might expect AlCl\textsubscript{3} to be at a disadvantage because of its larger size when compared with BCl\textsubscript{3}, and so it is interesting to see that AlCl\textsubscript{3} can overcome this by initiating bonding at a further distance ($\sim$1 \AA~more than BCl\textsubscript{3}) from the surface. Overall, GaCl\textsubscript{3}'s behavior is very similar to AlCl\textsubscript{3}, and the OOP distance of GaCl\textsubscript{3} closely follows AlCl\textsubscript{3} until it nears its final adsorbed configuration.

Figure \ref{fig_pathways}(c) shows the adsorption energy barriers for BCl\textsubscript{3}, AlCl\textsubscript{3}, and GaCl\textsubscript{3}, into a SDW of various ADTP surfaces. The dashed lines indicate the maximum adsorption energy along the pathway. We define this as the adsorption barrier for chemisorption as measured from the reference point of zero adsorption energy. (See Figure \ref{fig_pathways}(b).) In several cases, we observed a slightly favorable adsorption energy (local minimum of absolute value $<$ 0.1 eV) for the precursor above the surface due to physisorption. So the total barrier was computed as the difference between the maximum energy and the physisorbed energy as shown in Figure \ref{fig_pathways}(b). We report that on the bare Si surface all precursors adsorbed without barrier or with negligible barrier, which is shown in the Supporting Information document along with the entire set of reaction pathways for all precursor-resist combinations. BCl\textsubscript{3} shows a negligible barrier for the H resist case (0.001 eV). The Cl, Br, and I resists all result in total adsorption barriers of at least 0.28 eV for BCl\textsubscript{3}. These barriers increase with increasing resist atom size. The barrier is attributed to steric hindrances of the resist atoms, which coheres with the trend. Similarly, for AlCl\textsubscript{3} and GaCl\textsubscript{3}, the H resist is no obstacle at all, and the steric hindrances are most noticeable for the I case. When comparing between acceptor adsorbates, we see again that the BCl\textsubscript{3} barriers are greater than those for AlCl\textsubscript{3} and GaCl\textsubscript{3}. In the Supporting Information, we show that PH\textsubscript{3} adsorbed without barrier into all SDWs.

\begin{figure*}
\centering
\includegraphics[width=\textwidth]{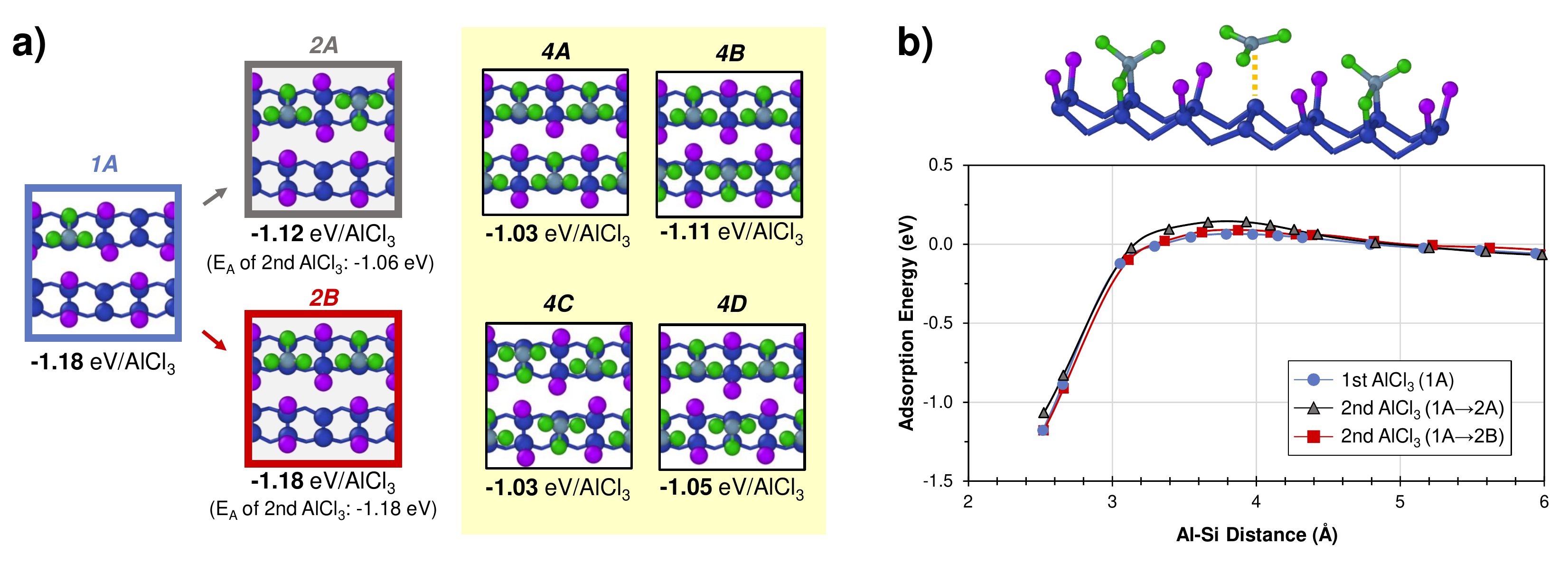}\hfill%
\caption{(a) AlCl\textsubscript{3}/I-ADTP models: single adsorbate model (1A), double adsorbate models (2A, 2B) representing filled adsorbate rows, and quadruple adsorbate models (4A, 4B, 4C, 4D) representing a dense adsorbate array where all SDWs are filled. Below each model is given the average AlCl\textsubscript{3} adsorption energy. (b) Adsorption reaction pathways for AlCl\textsubscript{3} into a SDW with and without neighboring occupied SDWs. The isolated adsorption case equates to the single AlCl\textsubscript{3} adsorption, ending in the 1A model. The neighboring adsorption case is given by introducing a second AlCl\textsubscript{3} to get to the 2A or 2B arrangement, and the reaction barrier depends on the orientation of the precursor relative to its neighbors.}%
\label{fig_array}%
\end{figure*}

The ease of SDW adsorption leads directly to the possibility of filling all windows of the ADTP. For the remainder of the paper, we consider adsorbate arrays, opting to focus on AlCl\textsubscript{3} arrays. The AlCl\textsubscript{3} array models are shown in Figure \ref{fig_array}(a). Also, shown is the single adsorbate case (1A) from Figure \ref{fig_ea} for reference. Since the presence of the adsorbate repels the I atoms, it brings to question whether the displaced I atoms would hinder adsorption into neighboring windows. To address this, first we consider a dimer row of completely filled SDWs. Two arrangements are shown in Figure \ref{fig_array}(a): one where the adsorbates alternate in orientation (2A) and one where the adsorbates are all oriented in the same direction (2B). Note that the boundaries are periodic. The adsorption energy for a single AlCl\textsubscript{3} on the I-ADTP surface is -1.18 eV. The dimer row is filled by a subsequent adsorption event, and the adsorption energy of the second AlCl\textsubscript{3} is computed by equation \ref{eq2}. Interestingly, when the orientation is the same, there is no additional energy penalty of the second adsorbate compared to the first. However, with alternating orientations the energy is less favorable by about 0.12 eV. This orientation preference is explored in the Supporting Information.

Going even further, quadruple adsorbate models (4A, 4B, 4C, 4D) were also calculated. These models represent a dense adsorbate array where all SDWs are filled. The average adsorption energy is about -1 eV per AlCl\textsubscript{3}, indicating that adsorption into the SDWs remains favorable. Four different adsorbate arrangements were modeled, and 4B was the most favorable of these. In model 4B, the adsorbates along the dimer row maintain the same orientation, but the orientations are flipped on the next dimer row. Regardless of how the adsorbates are oriented, the calculations suggest that it is possible to fill all SDWs forming a well-ordered array across the entirety of the surface.

While the arrays modeled in Figure \ref{fig_array}(a) appear viable based on the final energies, it behooves us to also examine the reaction pathway for inserting an adsorbate into a dense array. In Figure \ref{fig_array}(b), the reaction pathways are shown for adsorption into a SDW neighboring occupied SDWs. This was performed by modeling the adsorption step of the second AlCl\textsubscript{3} of the adsorbate rows in Figure \ref{fig_array}(a). Due to the periodic boundary conditions, the second AlCl\textsubscript{3} is sandwiched by already-adsorbed AlCl\textsubscript{3} molecules, as shown in Figure \ref{fig_array}(b). When adsorbing between occupied SDWs, the results show that  the energy barrier depends on the orientation of the AlCl\textsubscript{3} molecules relative to the neighboring adsorbates. When the orientation is the same, the total barrier is 0.16 eV, which is close to the barrier for isolated AlCl\textsubscript{3} adsorption into a SDW (0.13 eV). However, when the orientation is switched, adsorption is slightly more difficult, giving a total adsorption barrier of 0.25 eV. Even though there is some added difficulty with flipped orientations, this barrier is still quite small. Overall, the calculations show that a SDW neighboring occupied SDWs can be filled. Based on the results from Figure \ref{fig_array}, we do not anticipate any significant difficulties in subsequent adsorption steps leading to the occupation of all SDWs.

\section{Conclusion} \label{conclusion}

In summary, the computations presented here give merit to the use of guided self-assembly to achieve a large-scale, well-ordered array of dopants.  Using DFT calculations, we demonstrated the adsorption of dopant precursor molecules (PH\textsubscript{3}, BCl\textsubscript{3}, AlCl\textsubscript{3}, GaCl\textsubscript{3}) into SDW sites found on X-Si(100) (X = H, Cl, Br, and I) at 0.5 ML coverage.  The largest adsorption barrier of 0.34 eV for BCl\textsubscript{3} on the I-ADTP surface is surmountable under appropriate experimental conditions.  As such, our findings suggest that dopant precursor adsorption into SDWs should be feasible for all precursor and resist combinations studied here.  While this study demonstrates that self-assembled SDWs can foster a well-ordered array of adsorbed dopant precursors and gives merit to the use of self-assembled resist patterns, the ultimate success of this approach will depend on the ability to incorporate and electrically activate each dopant atom in place within the array.

Several strategies have been conceived that could be tested to incorporate and activate the dopants: a) standard thermal annealing, b) using the STM tip \cite{liu2016} or some other external stimuli (electrons, photons, or ions) to drive incorporation, or c) simply encapsulating both the adsorbed dopant precursors and the resist atoms in an epitaxial silicon capping layer followed by a rapid thermal anneal. A relatively straight forward experimental exploration is possible for each scenario.  In the near-term, any of the dopant precursors covered in this study can be easily tested with Br- or I-patterned Si(100) in combination with one of the above mentioned dopant  incorporation and activation strategies.  A standard Hall bar measurement of the resulting $\delta$-layer can be used to extract the active carrier concentration as well as the mobility for comparison to disordered $\delta$-layers fabricated using standard doping methods.  In particular, we expect the $\delta$-layer composed of a well-ordered dopant array to display a significantly higher mobility due to the highly ordered electrostatic potential within the transport layer.  We would also expect a higher carrier concentration as the resist layer will serve to prevent dopant pairing, which leads to the formation of electrically inactive dopant complexes \cite{campbell2021}.

If well-ordered dopant arrays in Si(100) can be achieved through guided self-assembly, it will then be worth investigating other self-assembled resist patterns with increased SDW spacing. For certain applications, such as spin-based quantum computing and quantum simulation, a dopant spacing $>$ 5 nm would be desirable.  With a view to that end, we note that Xu et al. reported a stable structure on I-Si(100) at 0.8 monolayer coverage, resulting in SDW spacings $>$ 1 nm.\cite{xu2005}

As solid-state quantum computers present a tremendous fabrication challenge, it demands that numerous approaches are brought to bear on the problem. We see a possibility for innovative strategies that leverage the halogens' innate chemistry, especially considering the current impediments to the use of STM for atomic precision patterning. Indeed, the concepts explored in this study are very much untapped but hold potential for considerable efficacy.

\section{Acknowledgement}

The authors acknowledge the computational facilities from the University of Maryland supercomputing resources and the Maryland Advanced Research Computing Center (MARCC).

\section{Supporting Information}

Rotation of a BCl\textsubscript{2} fragment, PH\textsubscript{3} undissociated adsorbate, and PH\textsubscript{2} fragment within a SDW. SDW adsorption reaction pathways for all precursor-resist combinations studied. Orientation preference for AlCl\textsubscript{3} adsorbates forming an adsorbate row.

\bibliography{window_paper}

\end{document}